\documentclass[9pt,twocolumn,twoside]{opticajnl}
\journal{opticajournal} 

\setboolean{shortarticle}{true}

\usepackage{lineno}

\title{Multiplexed Metasurfaces for Diffractive Optics via Phase Correlation Method}

\author[1,3]{Chenxuan Xiang}
\author[2,3]{Jumin Qiu}
\author[1]{Qiegen Liu}
\author[1]{Shuyuan Xiao}
\author[1,*]{Tingting Liu}

\affil[1]{School of Information Engineering, Nanchang University, Nanchang, Jiangxi 330031, China}
\affil[2]{School of Physics and Materials Science, Nanchang University, Nanchang, Jiangxi 330031, China}
\affil[3]{These authors contributed equally to this work.}
\affil[*]{ttliu@ncu.edu.cn}

\begin{abstract}
The multiplexing capability of metasurfaces has been successfully demonstrated in applications such as holography and diffractive neural networks. However, identifying a suitable structure that simultaneously satisfies the phase requirements across multiple channels remains a significant challenge in many multiplexing design scenarios. In this study, we propose an innovative phase correlation method for metasurface multiplexing design that utilizes a multi-layer perceptron to establish phase correlations across multiple channels. This approach reduces the difficulty of multi-channel phase training by converting it into a simpler single-channel optimization task, thereby reducing design complexity and computational cost. Using the proposed method, we design a dual-wavelength multiplexed diffractive neural network and a multi-wavelength metasurface color holography under a linear polarization. The designed multiplexed metasurface achieves up to 90\% classification accuracy in image recognition and exhibits good performance in color holography.

\end{abstract}

\setboolean{displaycopyright}{false} 

\begin{document}

\maketitle

Metasurfaces, as two-dimensional (2D) metamaterials, combine the unique properties of metamaterials while overcoming the limitations of traditional designs. In recent years, metasurfaces have emerged as a transformative platform for the design and fabrication of advanced optical elements and systems, demonstrating capabilities that surpass those of traditional diffractive optical components\cite{yu2014flat,sun2019electromagnetic}. The meta-atoms constituting metasurfaces can precisely control the response of incident electromagnetic waves\cite{yu2011light,shaltout2019spatiotemporal,liu2024arbitrarily}. These capabilities enable significant advancements in device miniaturization across various applications, such as vortex beam generators\cite{mei2023cascaded,tang2019high}, metasurface spectrometer\cite{wen2024metasurface,zhan2024single}, and polarization analyzer\cite{rubin2018polarization,guo2019high}. By manipulating physical dimensions such as phase, amplitude, and polarization, the powerful multiplexing capabilities of metasurfaces have been applied in holography\cite{arbabi2019vectorial,xu2024orthogonality,zhang2023high,huang2022orbital}, metalenses\cite{jiang2023multiwavelength,zhou2020helicity,min2024varifocal,lin2019achromatic}, and neural networks\cite{lin2018all,luo2022metasurface,lu2024metasurface,chi2024metasurface}. 

Metasurface multiplexing design strategies are generally categorized into forward design and inverse design. Forward design requires a deep understanding of physical principles to determine the arrangement of meta-atoms. For instance, in multi-wavelength metasurface color holography design, a direct approach involves embedding multiple nanostructures within each metasurface pixel, with each structure tailored to respond to a specific wavelength, enabling multi-wavelength modulation\cite{spagele2021multifunctional,song2020ptychography,georgi2021optical}. However, this method increases pixel size, which reduces display resolution and limits diffraction efficiency. These challenges can be addressed by employing meta-atoms with higher degrees of freedom\cite{chen2019broadband,overvig2019dielectric}, though this approach inevitably increases the complexity of the design. To resolve the inherent limitations of forward design, inverse design \cite{so2023multicolor,ma2022pushing,yin2024multi} has emerged as an innovative solution for metasurface multiplexing. For instance, Ref. \cite{so2023multicolor} optimizes a single phase profile using a gradient descent-based inverse design method to generate multiple holographic images. With advances in machine learning, end-to-end design \cite{ma2022pushing,yin2024multi} has been introduced to further optimize metasurface multiplexing. For instance, the end-to-end design framework proposed in Ref.\cite{yin2024multi} integrates the entire computational process into a differentiable framework, enabling performance enhancement through a single optimization step. Despite these advancements, challenges such as the complexity of constructing meta-atom libraries and long training times for optimization remain common in these design approaches. 

\begin{figure}[htbp]
    \centering
    \includegraphics[width=1\linewidth]{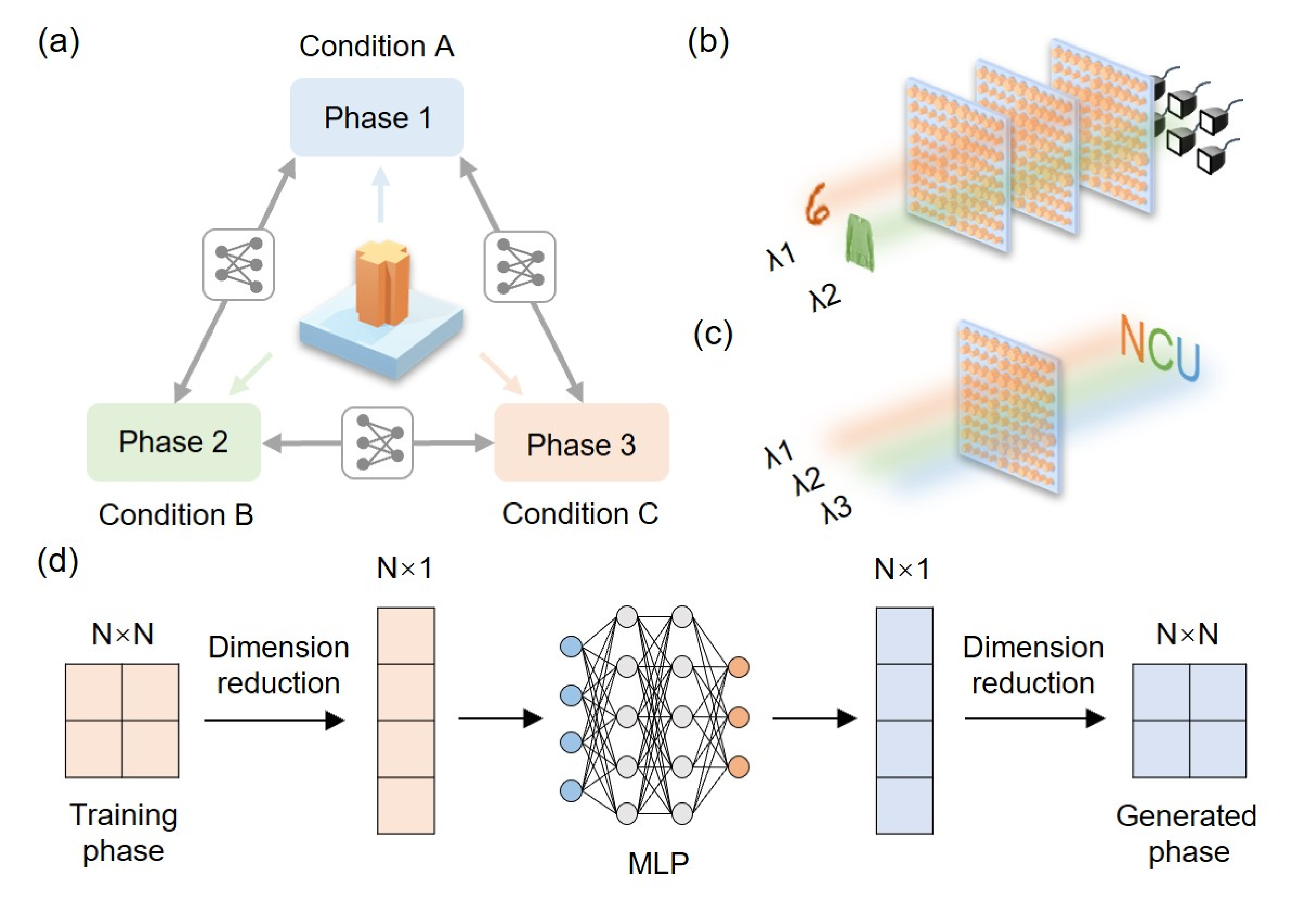}
    \captionsetup{justification=justified} 
    \caption{Principle of the phase correlation method and its applications in multiplexed metasurface design for diffractive optics. (a) The applicable scenarios of the phase correlation method. The metasurface generates different phases under different conditions, and the relationships between these phases can be modeled using the MLP. (b, c) Schematic illustration of dual-wavelength multiplexed diffractive neural network and multi-wavelength color holography. (d) Flowchart of the phase correlation method.
}
    \label{fig1}
\end{figure}

In this letter, we introduce the phase correlation method for multiplexed metasurface design. Under different conditions, the metasurface generates distinct phase responses. This method utilizes a neural network to correlate these phase responses, as shown in Fig. \ref{fig1}(a). As a concept of proof, the proposed method is demonstrated in two key applications: dual-wavelength multiplexed diffractive neural network (DW-MDNN) and multi-wavelength metasurface color holography (Fig. \ref{fig1}(b-c)) operating in the visible light. Both of the applications rely on a single meta-atom structure, significantly reducing the time required to build the meta-atom library. The phase correlation method uses parallel processing, which greatly reduces training and optimization time. By imposing appropriate constraints during phase training, we significantly reduce the difficulty of structure matching. 

The phase correlation method is central to the multiplexed metasurface design presented in this work. This approach correlates the phases of the metasurface under different conditions, thus reducing the complexity of structure selection. The method takes a training phase as input and employs a pre-trained multi-layer perceptron (MLP) as the fitting function. First, the training phase undergoes dimensionality reduction, and the reduced data is then fed into the pre-trained  MLP in parallel, which significantly accelerates the phase generation process. Finally, the output tensor from the MLP is expanded in dimensions to produce the generated phase, as shown in Fig. \ref{fig1}(d). 

\begin{figure}
    \centering
    \includegraphics[width=1\linewidth]{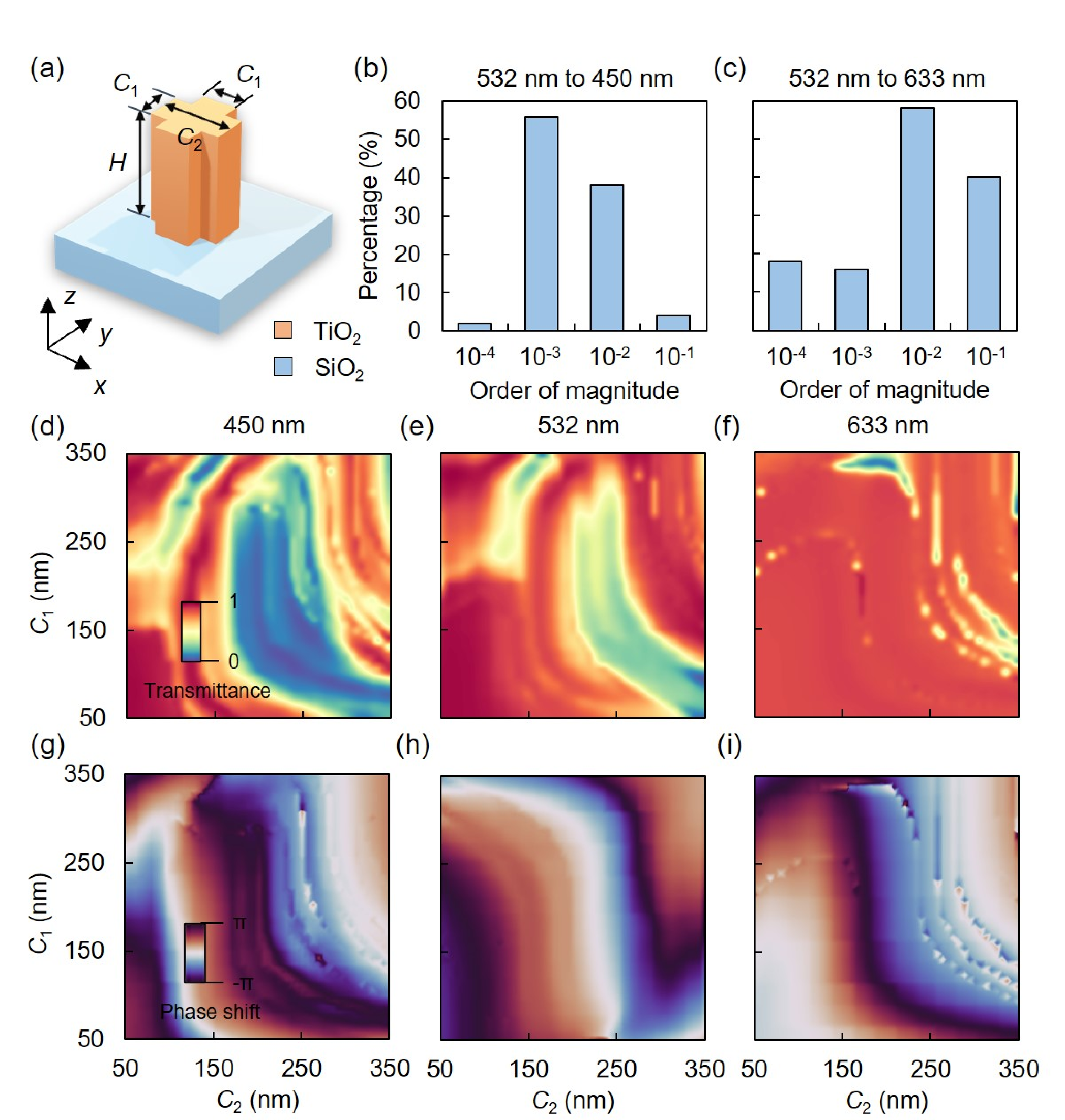}
    \captionsetup{justification=justified} 
    \caption{The design of the metasurface. (a) Schematic of a single TiO$_2$ meta-atom, where $H$ = 800 nm and $P$ = 400 nm (b, c) Fitting accuracy of the MLP model. (d-i) The simulated transmittance and phase shift of the meta-atom at wavelengths of 450, 532, and 633 nm, respectively.
}
    \label{fig2}
\end{figure}

The training data for the MLP is derived from finite difference time domain (FDTD) simulations. Fig. \ref{fig2}(a) illustrates the adopted meta-atom structure. The meta-atom is composed of subwavelength TiO$_2$ nanopillars with two structural parameters ($C1$, $C2$) and the height $H$ and the period $P$ are fixed. To demonstrate the phase correlation method, we employ a metasurface multiplexing design with three wavelengths: $\lambda_1$ = 450 nm, $\lambda_2$ = 532 nm, and $\lambda_3$ = 633 nm.  Fig. \ref{fig2}(d-i) show the simulated transmission response of each unit cell under $x$-polarized incident light for these wavelengths. In previous phase training without correlation, unconstrained training often increases the difficulty of selecting suitable structures. We construct two MLPs: one takes the phase at a wavelength of 532 nm as input and outputs the phase at 633 nm, while the other takes the phase at 532 nm as input and outputs the phase at 450 nm. Considering that the MLP will be involved in the training process, we design the MLP to have a simple network structure while maintaining low error, as shown in Fig. \ref{fig2}(b-c). Both MLPs consist of three hidden layers, each with 100 neurons, followed by a ReLU activation function. Here, we deliberately select only high-transmission structures, which significantly reduces the pool of available candidates in our database, and these data are then used to train the MLPs. 

After training the MLPs, we validate the application of the DW-MDNN in parallel multi-task classification. The MNIST and Fashion-MNIST datasets serve as inputs to the diffractive neural network.  Each image in the dataset is initially a 28 × 28 grayscale image, resized to 200 × 200, and then zero padded to 280 × 280. The MNIST datasets are endcoded at a wavelength of 532 nm, and the Fashion-MNIST datasets are endcoded at a wavelength of 633 nm, both under $x$-polarization. We use the phase correlation method to establish connections between the phases in different channels.  Although we consider the metasurface transmission factors, no correlation training is performed for them. We  train the phase in only one channel, while the phase in the other channel is automatically generated. This can greatly reduce the training time. The detection area is divided into four discrete rows, where the top two rows are used to predict the MNIST dataset, and the bottom two rows are used to predict the Fashion-MNIST dataset.

\begin{figure}[htbp]
    \centering
    \includegraphics[width=1\linewidth]{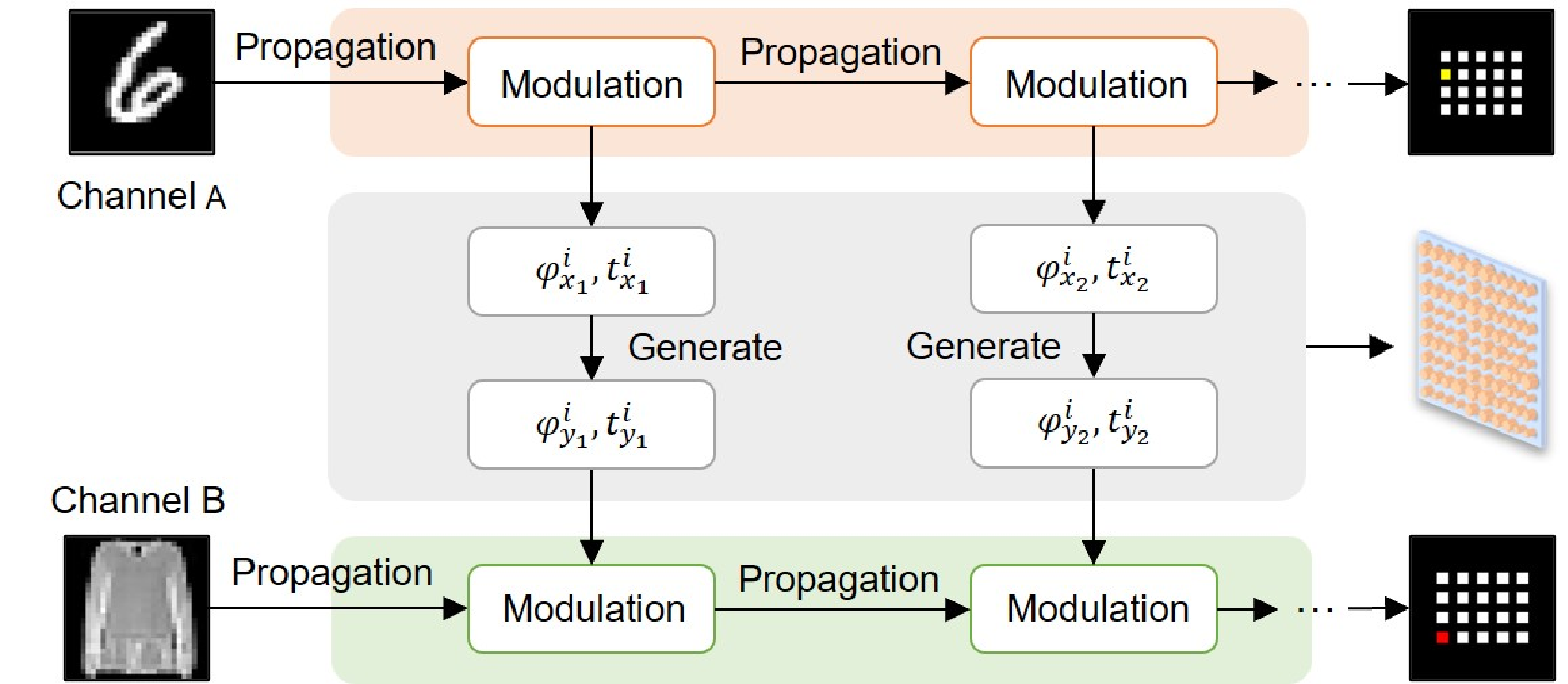}
    \captionsetup{justification=justified} 
    \caption{The flowchart of DW-MDNN. The MNIST and Fashion-MNIST datasets are encoded into two independent channels, A and B. Incident light of different wavelengths passing through the same metasurface generates different transmission coefficients and phase shifts. The phase correlation method is introduced to establish a relationship between the phases of two channels.
}
    \label{fig3}
\end{figure}

The diffractive neural network include  free-space propagation and metasurface modulation, as illustrated in Fig. \ref{fig3}. The method used for diffraction calculation in the free-space propagation is the angular spectrum method\cite{goodman2005introduction}, where the angular spectrum of the input light field can be expressed as $A_{0}\left(f_{x}, f_{y}\right)=\bar{F} E_{0}$, where $\bar{F}$  denotes the operator for the two-dimensional Fourier transform. The propagation of light from the input plane to the output plane at a distance $z$ can be described using the angular spectrum shift theorem as follows,
\begin{equation}
\begin{aligned}
A\left(f_{x}, f_{y}, z\right) = \exp \left[j \frac{2 \pi z}{\lambda} \sqrt{1-\cos ^{2} \alpha-\cos ^{2} \beta}\right] A_{0}\left(f_{x}, f_{y}\right),
\end{aligned}
\end{equation}
where  $f_{x}=\frac{\cos \alpha}{\lambda}$ and  $f_{y}=\frac{\cos \beta}{\lambda}$ represent the spatial frequencies of the planar light along the spatial axes $x$ and $y$ respectively. The output optical field on the $z$-plane can be obtained through the inverse Fourier transform as follows:  
\begin{equation}
\begin{aligned}E(z)=\hat{F}^{-1} \exp \left[j \frac{2 \pi z}{\lambda} \sqrt{1-\cos ^{2} \alpha-\cos ^{2} \beta}\right) \hat{F} E_{0}.
\end{aligned}
\end{equation}
The output optical field of the metasurface modulation can be expressed as $a_{x}(x, y) \cdot e^{j \varphi_{x}(x, y)}$, where $a_{x}(x, y)$ represents the amplitude of the metasurface, and $\varphi_{x}(x, y)$ represents the phase of the metasurface. Corresponding to different wavelengths, a specific meta-atom structure can produce varying phases and amplitudes. 

The phase correlation method is shared during training, so the loss function for network can be expressed as a weighted sum of the loss functions for the MNIST dataset and the Fashion-MNIST dataset. The formula is as follows, \begin{equation}
\begin{aligned}l_\text {total}=w_{1} * l_\text {MNIST}+w_{2} * l_{\text {Fashion-MNIST}}.\end{aligned}
\end{equation}
We set $w_{1}$ and $w_{2}$ at 0.41 and 0.59. Additionally, the Adam optimizer is used for gradient descent during the optimization process. This optimization algorithm combines the advantages of momentum and adaptive learning rates, enabling faster convergence while reducing oscillations near local minima, allowing the model to achieve optimal results within a shorter time frame.  

\begin{figure}[ht]
    \centering
    \includegraphics[width=1\linewidth]{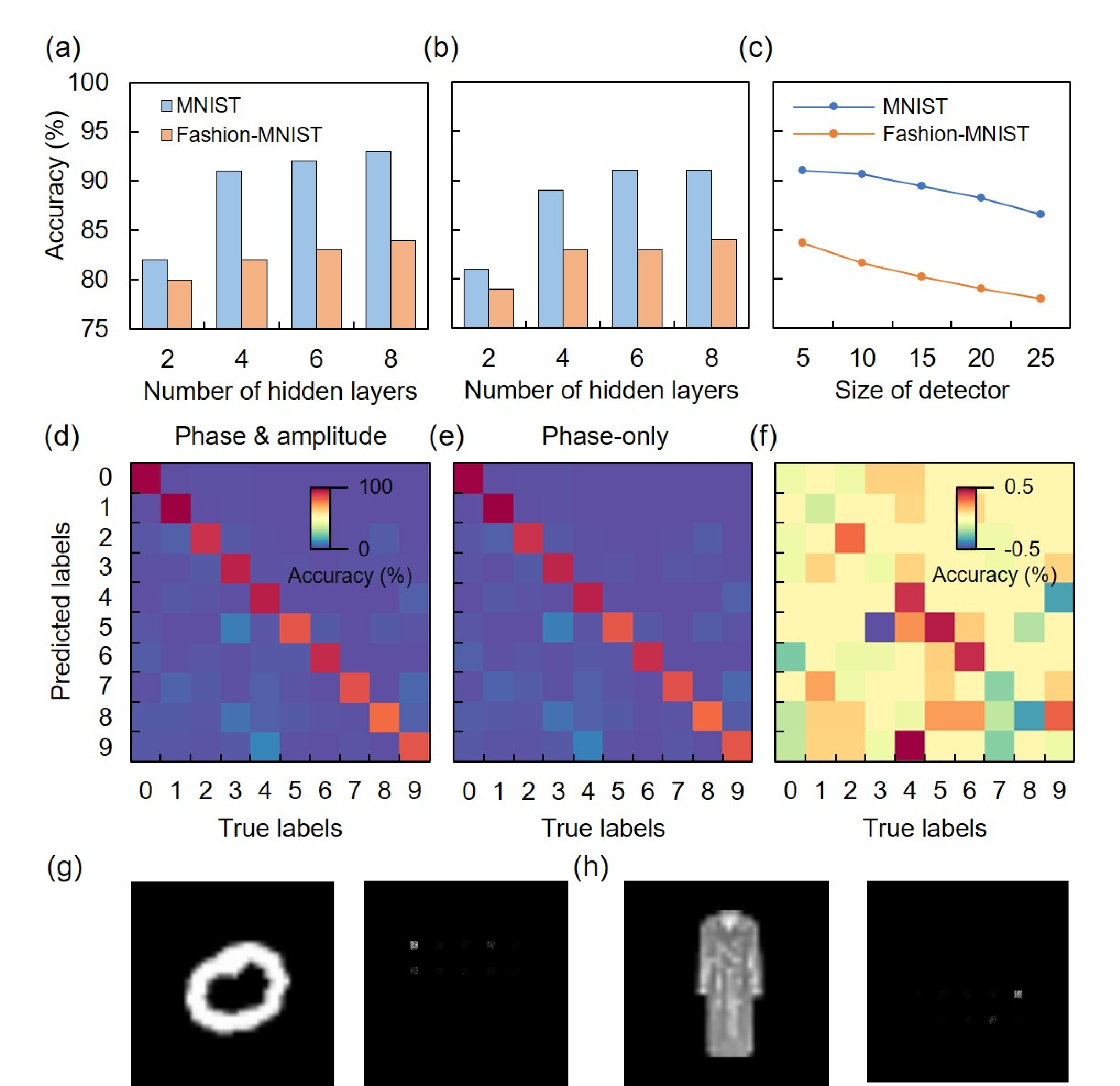}
    \captionsetup{justification=justified} 
    \caption{Results of DW-MDNN. (a, b) Classification accuracy as a function of the number of hidden layers without and with phase correlation. (c) Classification accuracy as a function of the size of the detector. (d, e) Confusion matrix for amplitude crosstalk and phase-only (amplitude is considered to be a constant value). (f) Percentage error matrix between (d) and (e). (g, h) Predicted results of the network.}
    \label{fig4}
\end{figure}

To validate the performance of our phase correlation method, we perform phase-uncorrelated training in Fig. \ref{fig4}(a) and phase-correlated training in Fig. \ref{fig4}(b) for DW-MDNN, respectively. In phase-uncorrelated training, we train the model individually on the MNIST and Fashion-MNIST datasets. To facilitate a better comparison, the hyperparameters for both training methods are set to be identical. The trained DW-MDNN is then tested on the test sets. The results indicate that the accuracy of phase-correlated training is comparable to that of phase-uncorrelated training (Fig. \ref{fig4}(a-b)), demonstrating that our approach significantly reduces the complexity of selecting meta-atom structures without compromising the accuracy of the diffractive neural network. Furthermore, the accuracy generally improves with an increasing number of hidden layers, suggesting that despite the lack of nonlinearity, the diffractive neural network still benefits from the "depth" advantage. As shown in Fig. \ref{fig4}(c), the detection size has a minimal impact on the accuracy of the diffractive neural network. Additionally, we include the amplitude factor of the metasurface in our analysis and observe that amplitude crosstalk has a negligible effect on the performance of the diffractive neural network (Fig. \ref{fig4}(d-f)). This is primarily because phase modulation plays a dominant role in the interaction between light and the metasurface. To further assess the network performance, we select one image from both the MNIST and Fashion-MNIST datasets as input. The resulting intensity distributions match the corresponding correct label outputs (Fig. \ref{fig4}(g-h)). We demonstrate the use of a five-layer network, where the phase correlation method theoretically achieves an accuracy of 90.78\% on the MNIST dataset and 82.88\% on the Fashion-MNIST dataset. The practical accuracy achieved after matching the metasurface structure is 90.68\% on the MNIST datasets and 82.68\% on the Fashion-MNIST datasets. This indicates that the error in structure matching using the phase correlation method is minimal. 

\begin{figure}[htbp]
    \centering
    \includegraphics[width=1\linewidth]{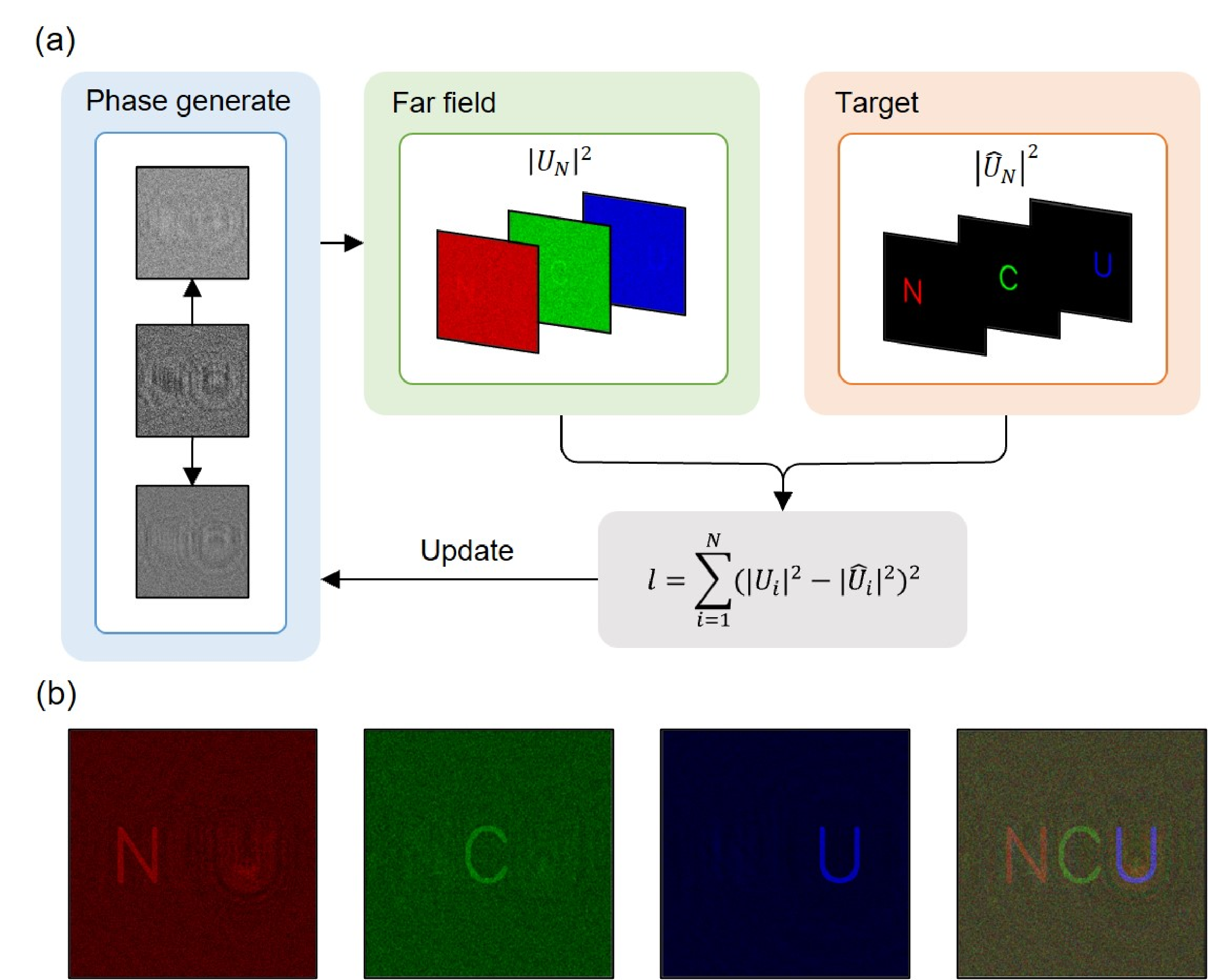}
    \caption{Flowchart (a) and results (b) of multi-wavelength color metasurface holography}
    \label{fig5}
\end{figure}

Further, to verify the feasibility of the design method in more channels, we implement a multi-wavelength color metasurface holography using the phase correlation method. The colorful "NCU" patterns are utilized as the target field, with images encoded into the corresponding channel's optical amplitude.  The forward computation of the holography still uses the angular spectrum method. Similarly, only the metasurface phase profile for one wavelength channel needs to be updated, while the phase profiles for the other two wavelength channels are automatically adjusted. This method utilizes a gradient descent algorithm, where the objective function is defined as the mean squared error (MSE) between the far field intensity $\left|U_{\mathrm{i}}\right|^{2}$ and the target field $\left|\widehat{U}_{i}\right|^{2}$. The metasurface phase distribution under wavelength $\lambda_2$ is treated as the optimization variable. The Adam optimizer is used to iteratively adjust the phase distribution to minimize the objective function, as shown in Fig. \ref{fig5}(a). We set the propagation distances to 1500 µm, and achieve excellent results, as shown in Fig. \ref{fig5}(b). 

In summary, we propose the phase correlation method for metasurface multiplexing design and demonstrate its application in the DW-MDNN and multi-wavelength color metasurface holography. The aim of this method is to address the challenge of finding a suitable structure that can simultaneously satisfy the phase requirements across multiple channels in metasurface multiplexing. By converting multi-channel training into single-channel training, the phase correlation method simplifies the design process. The phase correlation method can be applied to various diffractive optics scenarios, such as polarization multiplexing and wavelength multiplexing. Additionally, this method can be extended to other types of correlations, such as the complex amplitude correlation method.

\section*{Funding}
This work was supported by the National Natural Science Foundation of China (Grants No. 12364045, No. 12264028, and No. 12304420), the Natural Science Foundation of Jiangxi Province (Grants No. 20232BAB201040 and No. 20232BAB211025), and the Young Elite Scientists Sponsorship Program by JXAST (Grants No. 2023QT11 and 2025QT04)

\section*{Disclosures}
The authors declare no conflict of interest.

\section*{Data availability}
Data underlying the results presented in this paper are not publicly available at this time but may be obtained from the authors upon reasonable request.



\end{document}